 \documentclass[twocolumn]{webofc}

\usepackage[varg]{txfonts}   
\usepackage{hyperref}
\usepackage{url}
\usepackage{graphicx}

\usepackage[utf8]{inputenc}
\usepackage[T1]{fontenc}

\hypersetup{colorlinks=true,citecolor=blue,urlcolor=blue,linkcolor=blue}

\begin{document}

\title{Sensitivity of Isotopic Fission Yields in Actinides to the Macroscopic Liquid-Drop Model: LSD vs ISOLDA}

\author{\firstname{K.} \lastname{Pomorski}\inst{1} \and
        \firstname{A.} \lastname{Augustyn}\inst{1} \and
        \firstname{T.} \lastname{Cap}\inst{1} \and
        \firstname{Y. J.} \lastname{Chen}\inst{5} \and
        \firstname{M.} \lastname{Kowal}\inst{1}\fnsep\thanks{\email{m.kowal@ncbj.gov.pl}} \and
        \firstname{B.} \lastname{Nerlo-Pomorska}\inst{2} \and
        \firstname{M.} \lastname{Warda}\inst{2} \and
        \firstname{Z. G.} \lastname{Xiao}\inst{4}
}

\institute{National Centre for Nuclear Research, Pasteura 7, 02-093 Warsaw, Poland
\and
Department of Theoretical Physics, Maria Curie-Sk{\l}odowska University, 20-031 Lublin, Poland
\and
Department of Physics, Tsinghua University, Beijing 100084, China
\and
China Institute of Atomic Energy, Nuclear Data Center, Beijing 102413, China
}

\abstract{
The impact of the macroscopic liquid-drop prescription on isotope-resolved fission-fragment yields in the actinide region is assessed by comparing two alternative parameterizations: the Lublin--Strasbourg Drop (LSD) model and the ISOscalar Liquid Drop Approximation (ISOLDA). The two prescriptions differ primarily in the treatment of isospin dependence in the volume and surface terms; in ISOLDA, an explicit dependence on the isospin square $T(T+1)$, where $T=|N-Z|/2$, is introduced in both coefficients. Using an identical set of fragment-yield observables and the same experimental reference (fission of $^{250}$Cf$^*$ at low and high energies), the propagation of the macroscopic-energy choice into the predicted yields is quantified in terms of (i) the location of the most probable post-neutron isotopes along elemental chains, (ii) the widths and asymmetries of the isotopic distributions, and (iii) the population of neighboring nuclides on the distribution tails. A comparable description of the gross properties of the isotopic yield pattern is obtained with both prescriptions, particularly for light and intermediate fragments, where peak positions and near-maximum curvatures are reproduced similarly. The most discriminating differences are found for heavy-fragment chains, for which the ridge location and isotopic centroids are rendered more sensitive to macroscopic isospin terms. Overall, a closer average agreement with the evaluated data is obtained with LSD, while the LSD--ISOLDA spread is shown to provide a practical estimate of the macroscopic-model uncertainty in isotope-resolved yields.
}

\maketitle

\section{Introduction}
\label{sec:intro}
High-resolution isotopic fission-fragment yields, $Y(Z,N)$, are among the most challenging fission observables to measure experimentally. Their measurement requires an unambiguous, event-by-event determination of both the atomic number $Z$ and neutron number $N$ for fast-moving fragments, which severely limits the number of available datasets. As a result, only a small set of dedicated experiments has provided isotope-resolved yields with sufficient precision to constrain theory. In particular, inverse-kinematics measurements combined with high-resolution magnetic spectrometry at advanced facilities such as SOFIA \cite{Pellereau2017} and VAMOS \cite{Ramos2018,Ramos2019PRL239U} have opened the possibility of mapping detailed isotopic distributions over broad regions of the fission-fragment chart.

From the theoretical standpoint, reproducing isotope-resolved yields is equally demanding. A quantitative description is required to account simultaneously for the dissipative collective descent from saddle to scission and for the subsequent statistical de-excitation of the nascent fragments, while maintaining internal consistency among correlated observables such as mass and charge yields, total kinetic energy (TKE), and prompt-neutron multiplicity. In practice, the isotopic landscape provides one of the most stringent benchmarks for fission modeling because it probes not only the gross mass partition but also the fine balance between charge polarization, neutron sharing at scission, and post-scission evaporation.

To address these challenges, a wide spectrum of theoretical approaches has been developed, ranging from semi-empirical systematics to fully microscopic time-dependent frameworks. For instance, the GEF model \cite{Schmidt2016} has been shown to reproduce a broad range of global fission observables relevant to applications, whereas microscopic descriptions such as the Time-Dependent Generator Coordinate Method (TDGCM) \cite{Regnier2016,Verriere2021} provide a formally rigorous treatment rooted in effective nucleon--nucleon interactions, albeit at the price of substantial computational cost. In this context, stochastic transport models based on Langevin equations \cite{Aritomo2014,Usang2019} have emerged as a practical compromise, offering a physically motivated description of dissipative saddle-to-scission dynamics and a computational efficiency that makes systematic isotope-resolved studies feasible.

In the present work, isotope-resolved yields are calculated within a four-dimensional Langevin-plus-Master-Equation (4D LM) framework \cite{Pomorski2024LM}, in which the dissipative collective evolution is propagated in a multidimensional deformation space and the correlated post-scission de-excitation of primary fragments is treated statistically. The focus is placed on the sensitivity of the calculated isotopic patterns to the macroscopic sector of the underlying macroscopic--microscopic energy functional. Specifically, two alternative macroscopic prescriptions are considered: the Lublin--Strasbourg Drop (LSD) model \cite{PomorskiDudek2003LSD} and the iso-scalar liquid-drop approximation (ISOLDA) \cite{PomorskiXiao2025ISOLDA}.

The results obtained with the LSD-based implementation are highly encouraging: across a broad set of actinide cases, including spontaneous fission as well as neutron-induced fission at thermal and 14-MeV energies, the calculated isotope-resolved post-neutron yields reproduce the dominant ridge centroids and the near-maximum ``core'' shapes of the Ba and Xe chains with good fidelity~\cite{Pomorski2024LM,Pomorski2023PRC107}. This level of agreement provides strong validation of the present 4D LM framework and motivates a more stringent, systematic benchmark against evaluated experimental reference data compiled in the National Nuclear Data Center (NNDC) database at Brookhaven National Laboratory \cite{NNDC}.
At the same time, the comparison reveals a notable challenge~\cite{Pomorski2024LM}: for the highly excited $^{250}$Cf system ($E^{*}=46$~MeV, $L=20\hbar$) the agreement for isotopic yields deteriorates markedly, even though the corresponding mass distributions remain described reasonably well. This contrast suggests that the discrepancy is not simply a failure of the gross mass partition but rather reflects more subtle deficiencies in the isotope-forming stage. To help isolate the origin of these deviations, we therefore repeat the isotope-resolved calculations using the newly developed isoscalar liquid-drop model ISOLDA~\cite{PomorskiXiao2025ISOLDA} and perform a dedicated LSD--ISOLDA comparison, with the specific aim of quantifying the sensitivity to the macroscopic part and identifying the regimes where both macroscopic prescriptions converge versus those where the remaining shortcomings must originate elsewhere (e.g., in the fluctuation strength and its coupling to de-excitation).

\section{Methodology}
\label{sec:method}

The fission process is governed by a strongly dissipative large-amplitude collective motion in which the collective descent from saddle to scission is coupled to intrinsic degrees of freedom and accompanied by stochastic fluctuations. Within this physical picture, fragment observables emerge from an ensemble of trajectories whose drift is controlled by the thermodynamic driving potential, while their dispersion reflects the fluctuation--dissipation balance. In the present work, this dissipative dynamics is described within the four-dimensional Langevin-plus-Master-Equation framework, which enables a unified treatment of collective transport, pre-scission energy dissipation, and correlated post-scission de-excitation in actinide fission, including multi-chance fission following successive pre-fission neutron emissions \cite{Pomorski2024LM}.

Accordingly, the collective evolution is propagated by the multidimensional Langevin equations written for generalized collective coordinates $\vec q=\{q_i\}$ and conjugate momenta $\vec p=\{p_i\}$ as
\begin{equation}
\dot q_i=\sum_{j}\big[{\cal M}^{-1}(\vec q\,)\big]_{ij}\,p_j ,
\label{eq:langevin_q}
\end{equation}
\begin{equation}
\begin{aligned}
\dot p_i={}&-\frac{1}{2}\sum_{j,k}
\frac{\partial\big[{\cal M}^{-1}(\vec q\,)\big]_{jk}}{\partial q_i}\,p_jp_k
-\frac{\partial V(\vec q\,)}{\partial q_i} \\
&-\sum_{j,k}\gamma_{ij}(\vec q\,)\big[{\cal M}^{-1}(\vec q\,)\big]_{jk}\,p_k
+{\cal F}_i(t)\,,
\end{aligned}
\label{eq:langevin_p}
\end{equation}
where ${\cal M}(\vec q\,)$ denotes the collective inertia tensor, $\gamma(\vec q\,)$ the friction tensor, $V(\vec q\,)$ the thermodynamic driving potential, and ${\cal F}_i(t)$ a stochastic force. In the present implementation, ${\cal M}$ and $\gamma$ are evaluated within the irrotational-flow and wall-type approximations, respectively, following Ref.~\cite{Bartel2019CPC}. The fluctuation strength is coupled to dissipation in the standard manner, including the effective-temperature treatment of quantum fluctuations \cite{PomorskiHofmann1981}. The resulting set of stochastic differential equations is used to generate an ensemble of saddle-to-scission trajectories whose termination defines a scission-point event sample.

The generalized coordinate vector is taken as $\vec q=(c,a_3,a_4,\eta)$ and is defined within the Fourier-over-Spheroid (FoS) shape family \cite{Pomorski2023PRC107,PomorskiNerlo2023FoSNonax}. Here $c$ controls elongation, $a_3$ encodes left--right (mass) asymmetry, $a_4$ governs necking, and $\eta$ parametrizes non-axiality through elliptical transverse cross sections, thereby generalizing the role of the classical Bohr non-axiality parameter in standard collective models \cite{PomorskiNerlo2023FoSNonax}. The FoS representation provides a flexible description of strongly deformed, necked, and reflection-asymmetric configurations while preserving volume through the standard FoS constraint and maintaining the center of mass at the coordinate origin by an appropriate longitudinal shift; the explicit analytical expressions (profile function, mapping to non-axial shapes, and constraints) are adopted from Refs.~\cite{Pomorski2023PRC107,PomorskiNerlo2023FoSNonax} and are not repeated here.

The driving potential $V(\vec q\,)$ is constructed from a macroscopic--microscopic energy functional evaluated for FoS shapes \cite{Pomorski2023PRC107}. The macroscopic part is taken either from the Lublin--Strasbourg Drop (LSD) model \cite{PomorskiDudek2003LSD} or from the iso-scalar liquid-drop approximation (ISOLDA) \cite{PomorskiXiao2025ISOLDA}. Both methods and the differences between them are described further in the text. The microscopic shell and pairing corrections are obtained using a Yukawa-folded single-particle potential \cite{Dobrowolski2016YukawaFolded}. Finite-excitation effects generated by dissipation are incorporated through a Helmholtz-free-energy driving potential, built from the temperature-dependent macroscopic and microscopic contributions and a deformation-dependent level-density parameter. In particular, the quadratic temperature increase of the macroscopic sector and the progressive damping of microscopic corrections with temperature are implemented following Ref.~\cite{NerloPomorska2006PRC}. This thermodynamic construction provides a consistent measure of the collective drift at finite excitation without introducing additional ad hoc assumptions beyond the established LM prescriptions.

To assess the sensitivity of isotope-resolved fragment yields to dissipation, wall and microscopic friction prescriptions are considered together with the phenomenological temperature-dependent interpolation discussed in Ref.~\cite{IvanyukPomorski1996}. Scission-point observables are extracted from the ensemble of trajectories terminating at the scission criterion implemented in the FoS collective space, namely $a_4\approx 0.72$, which corresponds to a neck radius close to the size of a single nucleon~\cite{Pomorski2024LM}. In particular, the total kinetic energy (TKE) is evaluated from a scission-point energy balance that combines Coulomb repulsion between the nascent fragments, the nuclear interaction associated with neck rupture, and the collective pre-scission kinetic contribution, following the numerical prescription of Ref.~\cite{Pomorski2024LM}.

The pre-scission competition between fission and neutron emission is described with a set of coupled Langevin-plus-Master equations, similarly to what has been done in Refs. \cite{Pomorski2024LM,Pomorski2000NPA}, but using the FoS shape parametrization. The post-scission de-excitation of primary fragments is treated within the statistical model, through which event-by-event correlations among fragment properties are preserved \cite{Pomorski2024LM}. For the excitation-energy range relevant to the present benchmarks, charged-particle emission is strongly suppressed and neutron evaporation dominates \cite{Pomorski2000NPA}. Neutron-emission widths are evaluated within the Weisskopf--Ewing formalism \cite{Delagrange1986} using the Dostrovsky--Fraenkel--Friedlander parametrization for inverse cross sections \cite{Dostrovsky1959} and standard Fermi-gas level-density expressions with deformation-dependent level-density systematics taken from Ref.~\cite{NerloPomorska2002PRC}. In this way, the LM chain---from dissipative transport in the FoS-defined collective space to correlated statistical de-excitation---provides the basis for the isotope-resolved yield calculations and for the controlled comparison of macroscopic liquid-drop inputs (LSD versus ISOLDA) reported below.

\section{Results}
\label{sec:results}

As a benchmark for assessing the impact of the macroscopic energy prescription on isotope-resolved fission yields, we consider the fission of the excited $^{250}$Cf nucleus. Two cases are examined: low-energy fission, in which $^{250}$Cf is formed via thermal-neutron capture on $^{249}$Cf~\cite{NNDC}, and high-energy fission at $E^* = 46$ MeV, in which $^{250}$Cf is produced in the complete fusion of a $^{238}$U beam with a $^{12}$C target~\cite{PhysRevC.99.024615}. Particular attention is paid to the charge-assignment procedure at scission and to the differences between the two macroscopic prescriptions, LSD and ISOLDA, in their impact on the predicted isotopic yield patterns.

\subsection{Potential-energy landscape of \texorpdfstring{$^{250}$Cf}{250Cf}}
\label{sec:Cf250_pes}

\begin{figure}[h]
\centering
\includegraphics[scale=0.26]{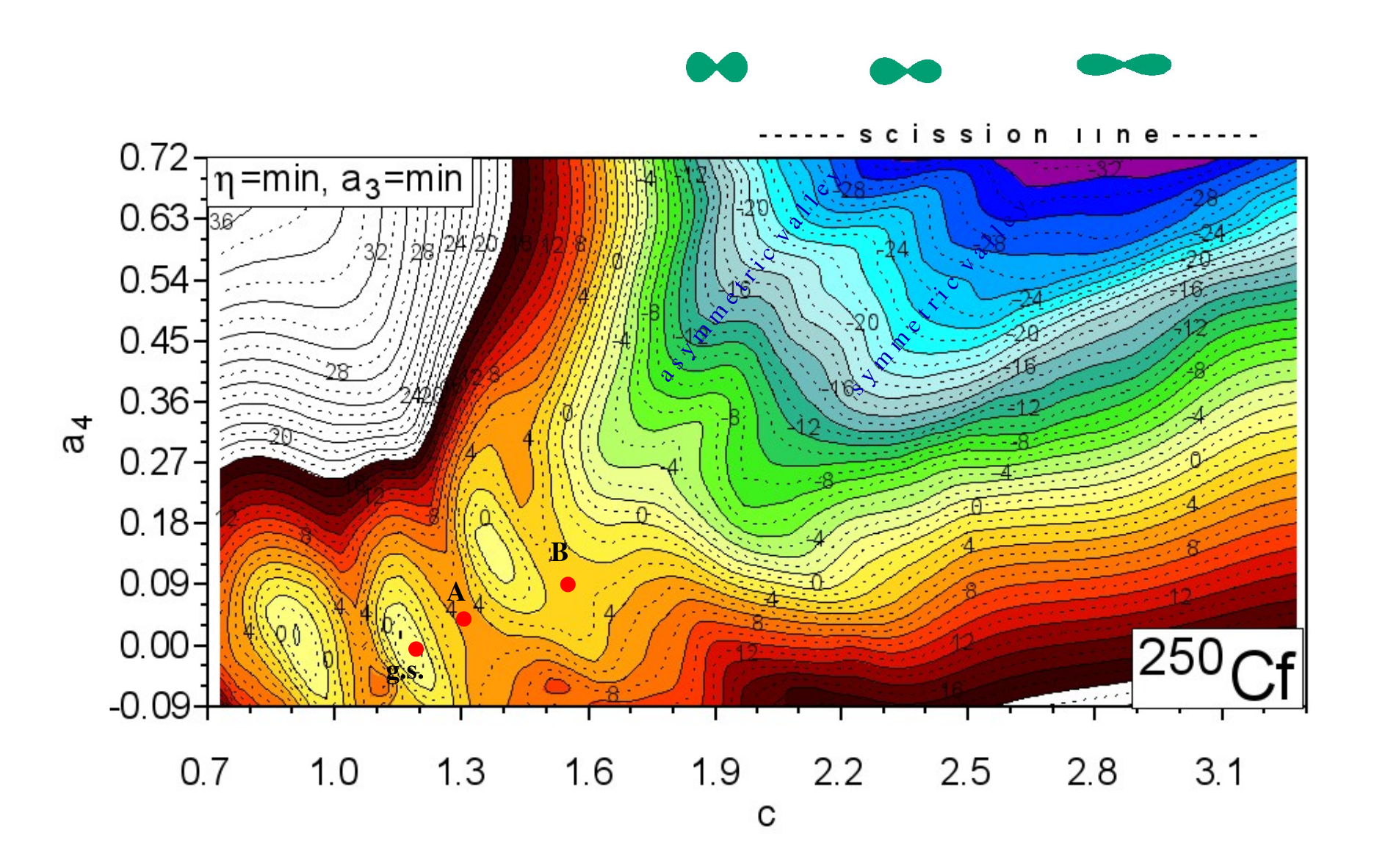}
\caption{Potential energy map of $^{250}$Cf in the $(c,a_4)$ plane, obtained by minimization over $a_3$ (mass asymmetry) and $\eta$ (nonaxiality). The label ``g.s.'' marks the first minimum; points $A$ and $B$ indicate reference configurations along a low-energy path; the dashed line denotes the approximate scission boundary. A separating ridge defines competing symmetric and asymmetric descent valleys.}
\label{fig:Cf250_pes}
\end{figure}

The minimized potential energy landscape of $^{250}$Cf obtained with the macroscopic-microscopic model (with the LSD macroscopic part) is shown in Fig.~\ref{fig:Cf250_pes} as a function of the elongation coordinate $c$ and the neck degree of freedom $a_4$. The surface displayed in the $(c,a_4)$ plane is obtained by minimizing the potential energy with respect to the remaining collective variables ($\eta=\min$ and $a_3=\min$). The plot is therefore to be understood as a two-dimensional projection of the full multidimensional landscape: while the detailed values of local barrier heights and the precise positions of stationary points may be affected by the minimization procedure, the global topology—minima, ridges, and valley connectivity that governs the stochastic flow—remains robust.

Several reference features are highlighted. The label ``g.s.'' indicates the vicinity of the first (ground-state) minimum.
The points $A$ and $B$ mark representative configurations along a low-energy pathway at increasing elongation and serve as convenient landmarks for the inner- and outer-barrier regions in the projected representation. The dashed line labeled ``scission line'' at $a_4\approx 0.72$ provides an approximate locus of scission configurations, beyond which a well-developed neck can no longer be sustained and the system evolves toward separated fragments.

Most importantly, two competing descent channels are resolved in the projected topology: a symmetric and an asymmetric fission valley, separated by a ridge. This bifurcation governs the partitioning of the probability flux between symmetric and asymmetric mass splits in subsequent Langevin dynamics and provides essential structural input for interpreting the calculated fragment-yield systematics.

Replacing the LSD macroscopic term with ISOLDA in the mass formula modifies the energy landscape; however, the overall conclusions and general characteristics remain as described above.

\subsection{Charge distribution of fission fragments}
\label{sec:charge}

The charge distributions of fission fragments are obtained as discrete yields over integer charge partitions at fixed isobaric splits, i.e.\ by assigning, for each scission event, a heavy-fragment charge $Z_h$ (and the complementary light-fragment charge $Z_l=Z-Z_h$) for a given heavy-fragment mass $A_h$ (with $A_l=A-A_h$). In order for such charge yields to be generated, a sampling probability must first be defined for each admissible integer charge partition. This is achieved by constructing Wigner-type weights that translate scission-energy differences between neighboring charge splits into a probability distribution,
\begin{equation}
W(Z_i)=\exp\left\{-\left[\frac{E(Z_i)-E_{\min}}{T^{*}}\right]^2\right\},
\label{eq:wigner_charge}
\end{equation}
which are then normalized,
\begin{equation}
P(Z_i)=\frac{W(Z_i)}{\sum\limits_j W(Z_j)}.
\label{eq:charge_prob_norm}
\end{equation}
A Monte Carlo draw is subsequently performed according to $P(Z_i)$ to select the heavy-fragment charge $Z_h$, while $Z_l$ follows from charge conservation. In this way, thermal scission-point fluctuations are directly propagated into the final charge yields.

To construct the weights of Eq.~\eqref{eq:wigner_charge}, the scission energy $E(Z_i)$ must be evaluated for each integer charge partition $Z_i$ that is compatible with the considered isobaric split. The effective temperature controlling the fluctuation scale is taken as
\begin{equation}
T^{*}=\frac{E_0}{\tanh(T/E_0)},
\label{eq:teff_charge}
\end{equation}
where $T$ denotes the thermodynamic temperature at scission and $E_0$ is chosen of the order of $\tfrac{1}{2}\hbar\omega_0$, with $\omega_0$ the harmonic-oscillator frequency. The quantity $E_{\min}$ in Eq.~\eqref{eq:wigner_charge} is defined as the minimum scission energy over the scanned set of integer charges for the given $(A_h,\vec q_h,\vec q_l)$, such that energy differences are measured relative to the most favorable partition. Here, $A_h$ is the mass number of the heavier fragment, and $\vec{q}_l$ and $\vec{q}_h$ denote the deformations of the nascent light and heavy fragments, respectively.

The evaluation of $E(Z_i)$ requires, for each scission configuration, the fragment deformations, as well as the scission configuration $\vec q_{\rm sc}$. With these inputs specified, the scission energy for a given heavy-fragment charge $Z_h$ is written as \cite{Pomorski2023PRC107}
\begin{equation}
\begin{aligned}
E(Z_h)={}&E_{\rm mac}(Z-Z_h,\,A-A_h;\vec q_l)
+E_{\rm mac}(Z_h,\,A_h;\vec q_h) \\
&+E_{\rm Coul}^{\rm rep}-E_{\rm mac}(Z,\,A;0)\,,
\end{aligned}
\label{eq:charge_energy}
\end{equation}
where $E_{\rm mac}$ denotes the deformation-dependent macroscopic energy evaluated within the LSD or ISOLDA model. The repulsive Coulomb interaction between the two deformed fragments at scission is taken as
\begin{equation}
E_{\rm Coul}^{\rm rep}=\frac{3e^2}{5r_0}\left[
\frac{Z^2}{A^{1/3}}B_{\rm Coul}(\vec q_{\rm sc})
-\frac{Z_h^2}{A_h^{1/3}}B_{\rm Coul}(\vec q_h)
-\frac{Z_l^2}{A_l^{1/3}}B_{\rm Coul}(\vec q_l)
\right],
\label{eq:coul_rep}
\end{equation}
with $r_0=1.217$~fm, consistent with the charge-radius constant.
The shape factors $B_{\rm Coul}$ account for the deformation dependence of the Coulomb self-energies and of the scission configuration.

Finally, in order for the sampling distribution to be anchored to the physically most probable charge split, the minimization of the scission energy with respect to the heavy-fragment charge is performed for each $(A_h,\vec q_h,\vec q_l)$. The integer charge $Z_h$ that minimizes Eq.~\eqref{eq:charge_energy} defines the most probable partition, while the set of neighboring integer charges provides the local energy landscape that enters Eqs.~\eqref{eq:wigner_charge}--\eqref{eq:charge_prob_norm}. The complete procedure is applied at the termination of each Langevin trajectory, thereby generating charge yields that consistently reflect both the scission energetics and the thermal fluctuations of the discrete charge degree of freedom.

\subsection{Sensitivity to the liquid-drop input: LSD versus ISOLDA}
\label{sec:lsd_isolda}

A key source of systematic uncertainty in macroscopic--microscopic fission modeling is the macroscopic liquid-drop input that shapes the large-scale topology of the potential energy surface and governs the mass partition between fragments at scission.

The macroscopic energy in the LSD model contains, in addition to the volume, surface, and Coulomb terms, a curvature and a congruence (Wigner) energy term:
\begin{equation}
\begin{split}
E_{\text{LSD}}(Z, A; \vec q_i) &= b_{\text{vol}} \left(1 - \kappa_{\text{vol}} I^2\right) A \\
&+ b_{\text{surf}} \left(1 - \kappa_{\text{surf}} I^2\right) A^{2/3} B_{\text{surf}}(\vec q_i) \\
&+ b_{\text{cur}} \left(1 - \kappa_{\text{cur}} I^2\right) A^{1/3} B_{\text{cur}}(\vec q_i) \\
&+ \frac{3}{5} \frac{e^2 Z^2}{r_0^{\text{ch}} A^{1/3}} B_{\text{Coul}}(\vec q_i) - C_4 \frac{Z^2}{A} \\
&+ E_{\text{cong}}(Z, N) + E_{\text{odd}}(Z,A).
\end{split}
\label{eq:lsd_ld}
\end{equation}
Here, $N$ is the neutron number, $b_{\rm vol}$, $b_{\rm surf}$, and $b_{\rm cur}$ (in MeV) are the volume, surface, and curvature coefficients. Corresponding $\kappa$ coefficients quantify $I$ dependence, where $I=(N-Z)/A$ is the reduced isospin. The fourth term is the Coulomb energy, and the fifth describes its reduction due to the surface diffuseness of the charge distribution, compared to a uniformly charged sharp-surface sphere, with $r_0^{\rm ch}$ being the charge radius parameter and $C_4$ the diffuseness coefficient. The last two terms describe congruence energy and the standard odd--even correction. The deformation dependence enters through the dimensionless shape factors $B_{\rm surf}(\vec q_i)$ (surface), $B_{\rm cur}(\vec q_i)$ (curvature), and $B_{\rm Coul}(\vec q_i)$ (Coulomb), normalized to unity for the spherical shape. The numerical parameter set and the explicit definitions of the shape factors are taken from Ref.~\cite{PomorskiDudek2003LSD}.

In the ISOLDA model the volume and the surface part of the macroscopic energy are dependent 
on the expectation value of the isospin square operator, $T(T+1)$, where $T=|N-Z|/2$:
\begin{equation}
\begin{split}
E_{\text{ISOLDA}}(Z, A; \vec q_i) &= -a_{\text{vol}} A \left(1 - 4k_{\text{vol}} \frac{T(T+1)}{A^2}\right) \\
&+ a_{\text{surf}} A^{2/3} \left(1 - 4k_{\text{surf}} \frac{T(T+1)}{A^2}\right)B_{\text{surf}}(\vec q_i) \\
&+ \frac{3}{5} e^2 \frac{Z(Z-1)}{r^{\rm ch}_0A^{1/3}}B_{\text{Coul}}(\vec q_i) \\
&- \frac{3}{4} e^2 \left(\frac{3}{2\pi}\right)^{2/3} \frac{Z^{4/3}}{r^{\rm ch}_0 A^{1/3}}\\
&+ E_{\text{odd}}(Z, A).
\end{split}
\label{eq:isolda_ld}
\end{equation}
Here, $a_{\rm vol}$, $a_{\rm surf}$, and $k_{\rm vol}$, $k_{\rm surf}$ have the same meaning as the analogous quantities in Eq.~\ref{eq:lsd_ld}, and the square of the elementary charge is $e^2\approx 1.44$ MeV$\times$fm. The numerical parameter set and the definitions of the shape factors are taken from Ref.~\cite{PomorskiXiao2025ISOLDA}. This additional isospin dependence modifies the driving potential and, consequently, the competition between neighboring isotopes along a given mass split. Note, the ISOLDA formula contains only 6 adjustable parameters, i.e., much fewer than the other commonly used macroscopic models.

\begin{figure*}[h]
\centering
\includegraphics[width=1.0\textwidth]{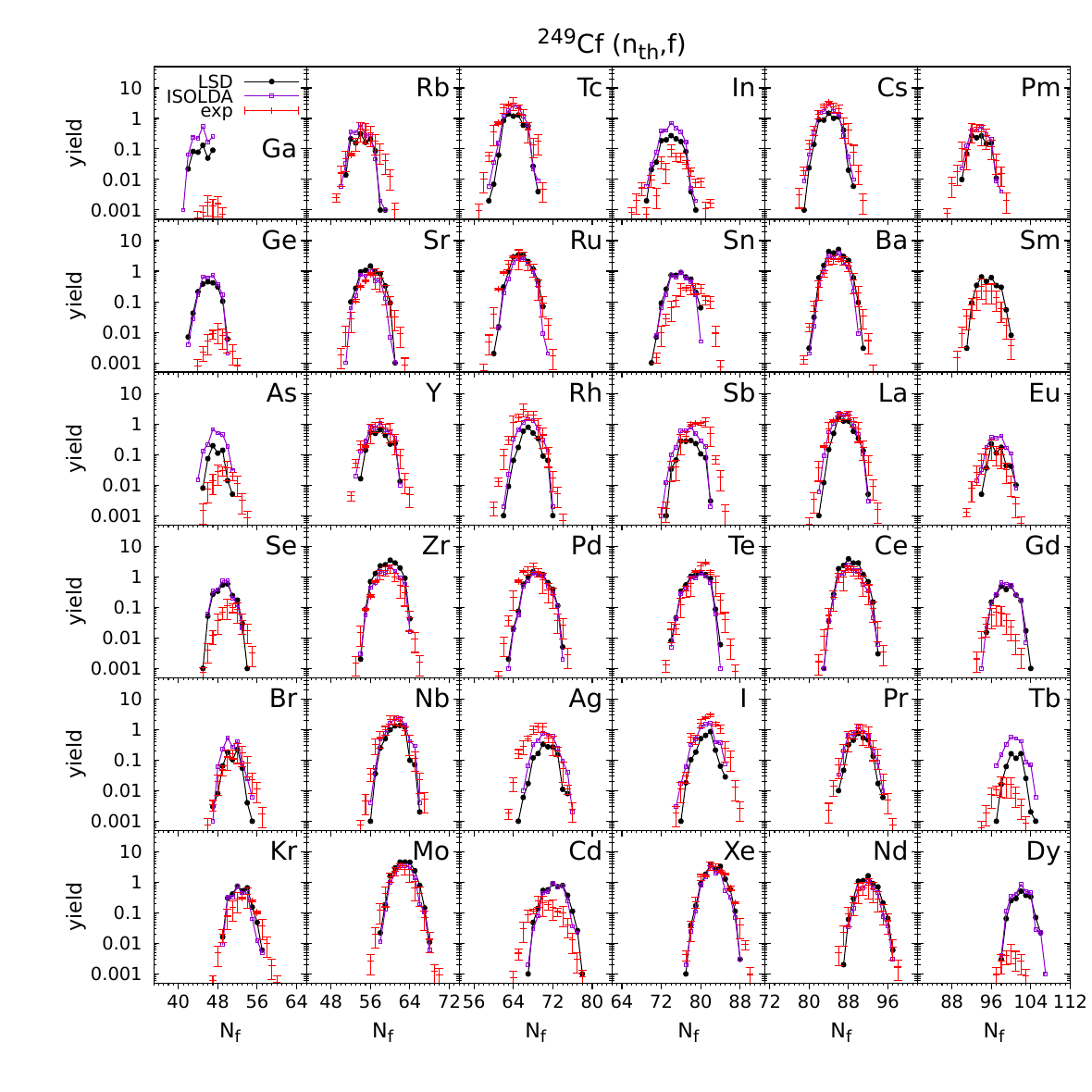}\hfill
\caption{Isotopic fission-fragment yields for $^{249}$Cf$(n_{\rm th},f)$ calculated within the 4D Langevin framework using the LSD  (points) and ISOLDA (squares) macroscopic prescriptions compared with evaluated experimental reference data (crosses) from~\cite{NNDC}. Each panel corresponds to a single elemental chain; the horizontal axis shows the post-neutron fragment neutron number $N_f$.}
\label{fig:cf250_nth}
\end{figure*}

In the low-energy fission case, both macroscopic prescriptions yield results in good agreement with experiment. As a representative example, Fig.~\ref{fig:cf250_nth} shows the calculated isotopic yields obtained with the LSD and ISOLDA prescriptions for the $^{249}$Cf$(n_\mathrm{th},f)$ reaction. The ISOLDA-based results are closely similar to the LSD ones. The isotopic fission-fragment yields are reproduced not only qualitatively but also quantitatively to a satisfactory degree. More notable discrepancies are observed only for weakly populated elemental chains, such as Ga, Ge, and Dy. In contrast, the chains corresponding to the most probable fragments, such as Ba and Xe, are reproduced with particularly high fidelity.

\begin{figure*}[h]
\centering
\includegraphics[width=1.0\textwidth]{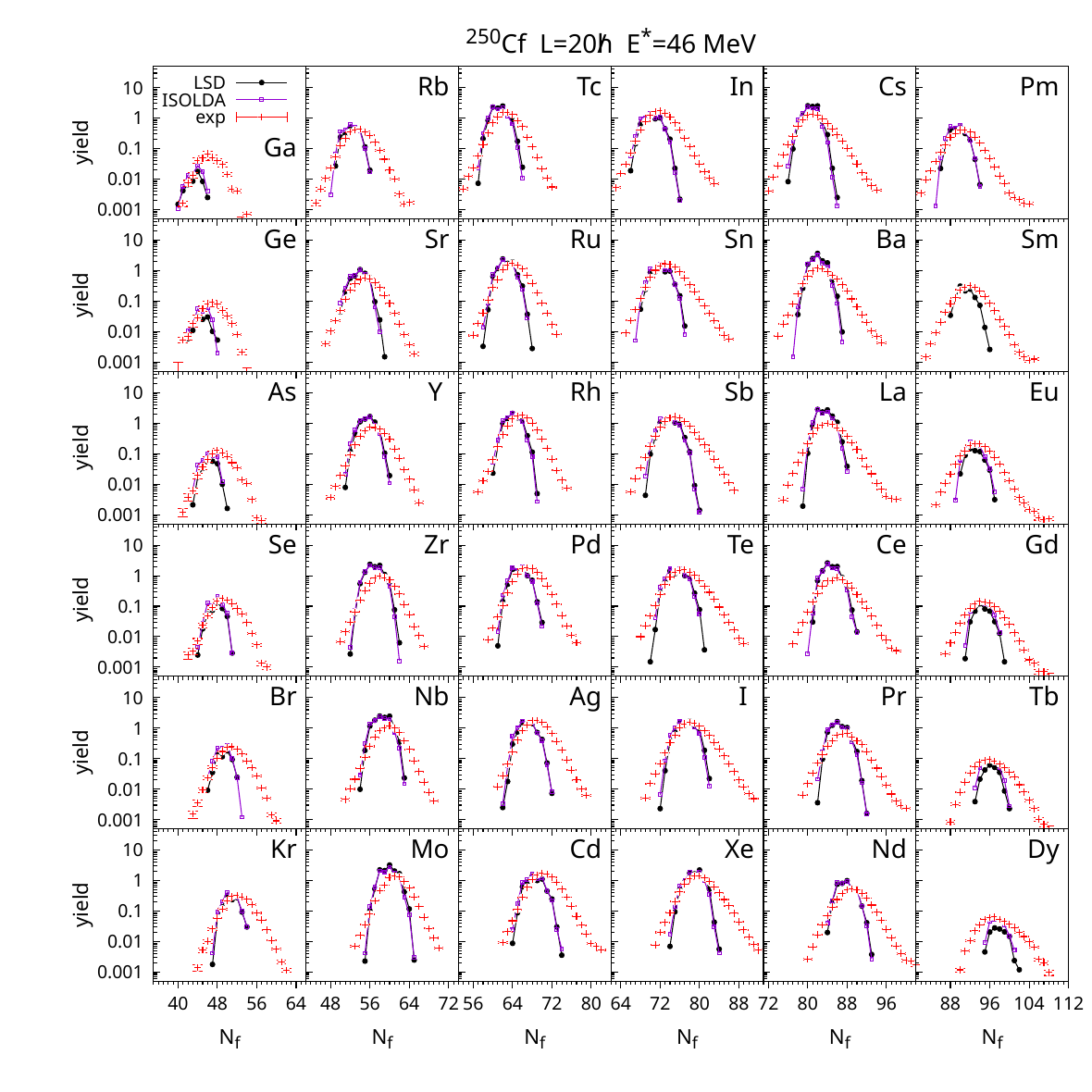}\hfill
\caption{Isotopic fission-fragment yields for $^{250}$Cf$^{*}$ ($L = 20\hbar$, $E^{*} = 46$ MeV) calculated within the 4D Langevin-plus-Master-Equation framework using the LSD (points) and ISOLDA (squares) macroscopic prescriptions compared with experimental data (crosses) from~\cite{PhysRevC.99.024615}. Each panel corresponds to a single elemental chain; the horizontal axis shows the post-neutron fragment neutron number $N_f$.}
\label{fig:cf250_ldlsd_compare}
\end{figure*}

Figure~\ref{fig:cf250_ldlsd_compare} compares the isotopic fragment yields obtained for high-energy fission of $^{250}$Cf produced in the complete fusion of a $^{238}$U beam with a $^{12}$C target~\cite{PhysRevC.99.024615}, at an excitation energy of $E^* = 46$ MeV and an average angular momentum of $L = 20\hbar$. These figures provide a stringent, isotope-resolved benchmark of the model performance by confronting calculated post-neutron isotopic yields with experimental distributions for a broad set of elemental chains. Since the two calculations differ only in the macroscopic liquid-drop prescription (LSD versus ISOLDA), the comparison isolates how the isospin dependence of the macroscopic energy propagates through the potential energy landscape and, ultimately, into the detailed isotopic pattern.

A first, robust observation is that \emph{both} macroscopic inputs reproduce the gross organization of the yield map: for many elements, the dominant ridge in $Y(N_f)$ is located in the correct neutron-number region, and the overall hierarchy of yields across neighboring isotopes is qualitatively captured. In particular, for several light and intermediate fragments (e.g.\ around Ga--Ge--As--Se and for parts of the Rb--Tc region), both LSD and ISOLDA deliver a satisfactory description of the peak location and the curvature in the vicinity of the maximum. This indicates that the underlying dynamical framework, together with the scission sampling and subsequent evaporation, captures the principal features of the mass partition and the associated charge polarization in this part of the fragment chart.

Beyond this common baseline, systematic differences between the two macroscopic prescriptions become apparent as one moves toward heavier fragment chains, where isotopic centroids and widths are especially sensitive to the balance among Coulomb, symmetry/isospin, and shell-driven polarization terms. In these chains (notably in the Cs--Ba--La--Ce--Nd and Xe region), the LSD-based calculation tends to track the experimental peak positions more closely, whereas the ISOLDA-based yields more frequently display an enhanced shift of the centroid and/or a distortion of the profile around the maximum. 
Given that the two calculations share the same dynamical treatment and the same scission-to-evaporation coupling, these discrepancies are naturally attributed to differences in the macroscopic driving potential, i.e.\ to how the respective liquid-drop functional constrains the energetics of partitions with different reduced isospin.

A second, equally important feature is that for a significant subset of elements both calculations tend to produce distributions that are too narrow when compared to experiment, most clearly in the heavy-fragment chains. The calculated curves often fall off faster on the distribution tails than the experimental data, indicating an underestimation of the event-by-event dispersion in neutron number at fixed $Z$. This behavior is largely \emph{model-independent} in the present comparison and therefore points to ingredients that are common to both runs: (i) the amount of effective stochasticity accumulated along the descent (controlled by the friction/diffusion balance), (ii) the degree of mixing between nearby descent channels in the multidimensional deformation space, and (iii) the strength of thermal fluctuations implemented in the discrete sampling of fragment charge and in the subsequent neutron evaporation. In other words, while the macroscopic input controls the centroid and the valley competition, the observed deficit in isotopic widths suggests that additional fluctuation strength and/or channel mixing may be required to fully reproduce the tails of the experimental distributions.

Taken together, the isotope-by-isotope assessment indicates that \emph{LSD provides the sightly better average agreement with experiment} for the present $^{250}$Cf benchmark. The advantage of LSD is most visible in the heavy-fragment sector, where it more reliably reproduces the position of the maxima and the overall shape near the peak, whereas ISOLDA tends to yield larger systematic deviations in the same region. The largest disagreements for both macroscopic inputs occur precisely where the isotopic systematics are most demanding---in the vicinity of the heavy-element chains such as Xe and Nd (and neighboring Cs/Ba), where both the centroid position and the breadth of the distribution are strongly affected by the interplay of the macroscopic isospin dependence, shell-driven polarization, and neutron evaporation. These shifts of the centriod are much larger than in the case $^{250}$Cf(n$_{\rm th}$,f) case which suggests that the post-neutron number is overestimated in our model. In this sense, these chains provide the most discriminating constraints on the macroscopic energy functional within a dynamical fission calculation.

In the preceding comparison, we have shown that several key isotope-resolved features are already reproduced with high fidelity. 
First, the locations of the neutron-number maxima $N_f$ in the Ba and Xe isotopic chains (the chain centroids) are captured remarkably well, with the best agreement obtained at lower excitation energies, including thermal-neutron-induced cases. 
Second, the shape of the distribution \emph{core} around the maximum---its local curvature and the relative population of neighboring isotopes---is described consistently, indicating that the mean charge partition and the average neutron content of the dominant fission channels are modeled in a coherent manner. 
Third, the heavy-light partner correlations are reproduced at least qualitatively: when the heavy-fragment maximum is correctly placed, the complementary light-fragment chain often follows with a comparably reasonable centroid, supporting the realism of the \emph{average} neutron sharing and the \emph{average} excitation-energy partition at scission.

The main remaining challenge is also clear and, importantly, it is \emph{common} to both macroscopic prescriptions. 
The calculated isotopic chains tend to be systematically too narrow, i.e.\ the yields fall off too rapidly on the tails, most notably for the heavy-fragment chains. 
This deficiency does not disappear when switching between LSD and ISOLDA: both macroscopic inputs lead to very similar isotope-resolved systematics, which implies that the width deficit is unlikely to be driven by the macroscopic term $E_{\rm mac}$ alone. 
In this sense, the role of the LSD--ISOLDA comparison is twofold: it demonstrates that the macroscopic sector is not the limiting ingredient for the present observable, and it provides a small but useful ``systematic spread'' that can be interpreted as an uncertainty band associated with the macroscopic energy, while the dominant physics to be improved must lie elsewhere.

Within the scope of the present framework and in line with the conclusions of this work, the most natural interpretation is that the common LSD/ISOLDA underestimation of isotopic widths in high-energy fission points to a slightly insufficient fluctuation content in the modeling chain, arising from a combination of (i) diffusion accumulated during the dissipative descent (effective friction strength $\gamma$ and associated stochastic driving), (ii) dispersion of scission-point conditions (thermal sampling of discrete fragment charge and excitation-energy sharing), and (iii) the stochasticity of prompt-neutron evaporation (including level-density inputs and their coupling to fragment deformation). 
Natural next steps are therefore systematic sensitivity studies of the fluctuation influence---by varying the effective dissipation strength, refining the thermal sampling of discrete fragment charges at scission, and improving the coupling to prompt-neutron evaporation (including excitation-energy sharing and level-density inputs)---with the specific goal of broadening the calculated isotopic distributions while preserving the accurately reproduced ridge centroids.

\section{Conclusions and outlook}
\label{sec:conclusions}
A Fourier-over-Spheroid (FoS) shape parametrization was employed as an efficient and flexible description of fissioning nuclear shapes, providing consistent coverage from the vicinity of the ground-state basin to strongly necked configurations near scission. On this basis, a macroscopic--microscopic potential energy landscape was constructed in the four-dimensional collective space $(c,a_3,a_4,\eta)$, combining a macroscopic liquid-drop term with microscopic shell and pairing corrections. Two alternative macroscopic energy prescriptions, LSD and ISOLDA, have been compared.

The fission process was described within a 4D Langevin approach, supplemented by a master equation for correlated pre-scission neutron emission, and by a statistical model for post-scission fragment cooling. 
The sharing of neutrons, protons, and excitation energy between nascent fragments at scission was treated consistently, and the resulting primary and secondary fragment yields in mass and charge were obtained. 

 A similar reproduction of the global structure of isotopic yield maps in high energy fission was obtained with both prescriptions, with the best agreement found for light and intermediate fragments. The largest discrepancies were observed for heavy-fragment chains, for which the ridge location and isotopic centroids were found to exhibit increased sensitivity to macroscopic isospin-dependent terms. On average, a slightly closer agreement with the reference data was obtained with LSD, while the LSD--ISOLDA spread was adopted as a practical estimate of the macroscopic-model uncertainty in isotope-resolved yields.

As a natural outlook, an extension of the collective space by incorporating higher-order Fourier deformations (5D FoS) is planned within the Langevin framework. An explicit treatment of the compound-nucleus angular-momentum distribution and its propagation into fragment observables is also intended, together with a refined inclusion of fragment shell effects. These developments are expected to improve the predictive power for isotope-resolved yields and to strengthen the description of their excitation-energy dependence.
\section*{Acknowledgments}

The authors (K.~Pomorski, B.~Nerlo-Pomorska, T.~Cap, M.~Kowal, and A.~Augustyn) gratefully acknowledge the Joint Research Centre (JRC) for support enabling their participation in the Theory-6 Scientific Workshop on ``Nuclear Fission Dynamics and the Emission of Prompt Neutrons and Gamma Rays''.

This research was funded in part by the National Science Centre, Poland, under Project No.~2023/49/B/ST2/01294. Additional support was provided by the Natural Science Foundation of China under Grant Nos.~11961131010 and 12335008. Furthermore, M.~Kowal and T.~Cap were partially supported by the Polish-French cooperation COPIGAL.


\end{document}